\documentclass[prc, groupedaddress, twocolumn, nofootinbib]{revtex4-1}

\usepackage{graphicx}
\usepackage{amsmath}
\usepackage{multirow}
\usepackage{enumerate}
\renewcommand\Im{\operatorname{Im}}
\usepackage{srcltx}
\usepackage{ulem}


\begin{document}
\title{Multinucleon ejection model for two-body current neutrino interactions}

\author{Jan T. Sobczyk}
\email{On leave from the Wroclaw University; e-mail: jsobczyk@fnal.gov}
\affiliation{{ }\\
Fermi National Accelerator Laboratory, Batavia, Illinois 60510, USA}


\begin{abstract}
A model is proposed to describe nucleons ejected from a nucleus as a result of two-body current 
neutrino interactions. The model can be easily implemented in Monte Carlo neutrino event generators. Various possibilities to 
measure the two-body current contribution  are discussed. The model can help 
identify genuine charge current quasielastic events and allow for a better determination of the systematic error on
neutrino energy reconstruction in neutrino oscillation experiments. 
\end{abstract}

\maketitle

\section{\label{sec:level1}Introduction}

There is a lot of evidence for a significant multinucleon ejection contribution to the inclusive neutrino charge current (CC) 
cross section in the 1~GeV energy region \cite{ccqe}. 
On the experimental side, several recent nuclear target CCQE (CC quasi-elastic) cross section measurements 
reported a large value for the 
axial mass ($M_A$), in disagreement both with older deuterium target measurements \cite{bodek_MA} and also with electroproduction 
arguments \cite{pion_MA}. 
A possible explanation for the discrepancy is that some events interpreted as CCQE are in fact due to a 
different dynamical mechanism
that typically leads to multinucleon emission \cite{martini1}. In fact, in the case of the MiniBooNE (MB)
CCQE measurement, nucleons in the final state were not analyzed at all. 
The MB collaboration reported a high-statistics 2-dimensional muon inclusive cross section likely to contain a large multinucleon
ejection contribution, and thus provide a challenge for theoretical models \cite{MB_MA}.

On the theoretical side there are several two-body current computations that support the idea of a 
significant multinucleon contribution to CC
neutrino scattering. 
First was the Marteau model, based on the earlier works of Ericson and Delorme \cite{marteau}. 
The model is formulated in the framework of the non-relativistic Fermi Gas in the Local Density Approximation (LDA) approach. 
It includes the QE and $\Delta$ excitation 
elementary interactions, Random Phase Approximation effects 
(RPA are in medium polarization corrections important in the low four-momentum transfer 
region \cite{luis}),
and $\Delta$ self-energy in the nuclear matter \cite{oset_salcedo}. 
The model was later upgraded by 
Martini, Ericson, Chanfray and Marteau (MEChM model) and compared to the MB data \cite{martini1}. 
It predicts a large np-nh contribution to the CC cross section that
can explain the MB CCQE anomalous $M_A$ measurement (n particles and 
n holes; in the Fermi Gas picture ejection of n particles means that there are also n holes left in the Fermi sea).
With the inclusion of relativistic corrections the MB 2D differential 
cross section can be reproduced \cite{martini2}. 

Another approach to the multinucleon ejection was proposed by Nieves, Ruiz Simo and Vicente Vacas \cite{nieves1}. 
The model uses techniques developed for the analysis of the 
inclusive electron-nucleus cross section in the kinematical region contaning both QE and $\Delta$ excitation peaks 
\cite{gil}. The model is relativistic and incorporates similar nuclear effects as the MEChM model.
In the paper \cite{nieves2} a comparison to the MB 2D CCQE data gave a best fit axial mass value $1.08 \pm 0.03$~GeV.

The third microscopic computation of the multinucleon ejection in electron and neutrino interactions was discussed in a series of  papers 
\cite{amaro}. There are many similarities but also some differences between the three approaches and a
detailed discussion of them may be found in \cite{comparison}. 

Recently, an effective approach to describe multinucleon contribution to the neutrino inclusive cross section was also
proposed.
The Transverse Enhancement model (TEM) \cite{bodek} is based on analysis of electron-carbon scattering data and 
parameterizes the multinucleon ejection contribution to the muon inclusive cross section in terms of a 
modification of the magnetic electromagnetic form factor. The model predicts that the two-body current contribution 
is less important at 
larger neutrino energies which can reduce a tension between the MB and NOMAD \cite{nomad} axial mass measurements. 

In all the approaches the calculated quantity is the contribution to the muon inclusive differential cross section corresponding to 
multinucleon ejection. 
This is sufficient if the aim is to 
reproduce the MB 2D
CCQE data. However, any attempt to separate the dynamical mechanism leading to the CCQE-like events (defined here as those with no pions in the 
final state)
must be based on a careful investigation of the hadrons in the final state. This introduces a new ingredient 
to the discussion, Final State
Interactions (FSI) effects. There are always CCQE-like events which are not really CCQE due to FSI effects. 
The major contribution comes from the
pion absorption but there are also events with several nucleons in the final state originating from a 
primary CCQE interaction with an energetic 
outgoing proton.
A further clarification is necessary: in this paper the term  {\it CCQE} will always refer to primary interactions. The basic picture is that 
of a nucleus as composed from quasi-free nucleons. 

In many neutrino oscillation experiments neutrino energy is reconstructed based on the muon information only.
It is clear that 
the two-body current contribution can introduce a significant bias and 
should be considered in the evaluation of systematic errors. 
It is important to develop models of the two-body current contribution which can be implemented in neutrino Monte Carlo (MC) event 
generators \cite{MC},
and for that one needs predictions for nucleons in the final state.
The aim of this paper is to propose a model to provide this information. The nucleon ejection part of the model is quite universal and can be 
used together with any model of the double differential muon inclusive cross section.

The paper is organized in the following way. In Sect. 2 the multinucleon ejection model is introduced. In order
to make numerical predictions a model of the 
muon 2D differential cross section is necessary. Two such models are described: a microscopic
model inspired by the Marteau approach and the effective TE model. Both models have been implemented in the 
NuWro MC event generator \cite{nuwro}. In Sect. 3 some predictions from the model are shown and the focus is on demostration of the 
relevance of the FSI effects. Section 4 contains a discussion of possible ways to measure the two-body contribution experimentally and 
a few final remarks can be found in Sect. 5.
%
\section{The model}
%
The multinucleon ejection model proposed in this paper is quite general. It 
needs as an input a muon inclusive differential cross section model. Two models
used in the numerical computations will be described in \ref{inclu_models}.
%
\subsection{Nucleon ejection model\label{model}}
%
One basic assumption of the model is that the energy and momentum are transfered to two (or three) nucleons simultaneously.  

The procedure to generate nucleon final states is as follows:

\begin{itemize}
\item Two (or three) nucleons are selected from a Fermi sphere of radius $220$~MeV 
(we assume the interaction occurs on carbon).
\item Th four momentum of the hadronic system is calculated by adding the four momenta of selected nucleons and the energy and momentum transfered
by the interacting neutrino.
\item A Lorentz boost to the hadronic center of mass system is performed.
\item Two (or three) nucleons are selected isotropically in the hadronic center of mass system
\item The boost back to the laboratory frame is performed.
\end{itemize}

Each event is weighted by the muon inclusive differential cross section. The
energy balance is done based on the assumption that inital state nucleons are in the potential well of the depth $V=E_f + 8$~MeV ($E_f$ is
the Fermi energy). In the numerical computations:

\begin{itemize}
\item The Fermi energy is subtracted from each initial state nucleon.
\item For each nucleon in the final state (in the LAB frame) its energy is reduced by $8$~MeV, adjusting the momentum
so that it remains on-shell.
\end{itemize}

The above procedure allows for a smooth distribution of the nucleon momenta in the final state. 
When the model is implemented in a MC event generator, after the initial interaction nucleons propagating
through a nucleus are subject to
rescatterings. The treatment of energy balance must be consistent with the way the cascade model is designed.

The algorithm introduces some correlations between the initial state nucleons: 
not all the initial
configurations give rise to a hadronic system with center of mass invariant mass larger than $2M$ (or $3M$ for 
three nucleon ejection, where $M$ is nucleon's mass). In the numerical code initial state nucleon configurations are selected until
acceptable nucleon momenta are found.

Tests were also done with fully correlated pairs of nucleons in the
deuteron-like configuration with opposite three-momenta in the initial state and the results remained virtually unchanged. The most important
assumption of the model is that of an isotropic distribution of final state nucleons in the hadronic center of mass frame.

In the Monte Carlo implementation it is assumed that 
in the CC neutrino reactions $\sim 80$\% of the final state nucleons are protons and only $\sim 20$\% are neutrons. For the 
CC antineutrino reactions
the isospin composition of the final state is opposite with $\sim 80$\% of neutrons and $\sim 20$\% of protons.
The authors of the MEChM model claim that in the case of CC
neutrino interactions there should be much more proton-neutron than neutron-neutron pairs in the initial state 
and as a consequence
more proton-proton than proton-neutron pairs in the final state \cite{martini1}. 

We note the above procedure to get nucleons in the final state is similar to the one considered earlier 
in \cite{oset_cascade}.
%
\subsection{Muon cross section models}
\label{inclu_models}
%
In this subsection two muon inclusive cross section models are presented.
%
\subsubsection{Microscopic model}
%
The model is based on the Marteau approach \cite{marteau_thesis}. It was developed about 10 years ago and discussed at the NuInt02 workshop 
\cite{sobczyk_nuint02}. The cross section is given as:
\begin{equation}
 \frac{d^2\sigma}{dq d\omega} = \frac{G_F\cos^2\theta_C q}{32\pi E_\nu^2} L_{\mu\nu}H^{\mu\nu},
\label{section}
\end{equation}
where $E_\nu$ is the neutrino energy, $q^\mu=(\omega, \vec q)$ is the four-momentum transfer, $q=|\vec q|$, $L_{\mu\nu}$ is the leptonic tensor 
\begin{equation}
 L_{\mu\nu} = 8\left( k'_\mu k_\nu + k_\mu k'_\nu -g_{\mu\nu}k\cdot k' \pm i\epsilon_{\mu\nu\alpha\beta}k'^\alpha k^\beta \right),
\label{leptonic}
\end{equation}
$k$ and $k'$ are neutrino and charged lepton four-momenta, $\pm$ signs refer to the neutrino and antineutrino cases.

The hadronic current is:
\begin{equation}
 H^{\mu\nu} = H^{\mu\nu}_{NN} + H^{\mu\nu}_{N\Delta} + H^{\mu\nu}_{\Delta N} + H^{\mu\nu}_{\Delta\Delta},
\end{equation}
where $H^{\mu\nu}_{XY}$, $X,Y\in (N,\Delta)$ are expressed in terms of expectation values of spin operators acting in the 
nucleus Hilbert space, so called nuclear response functions. The model uses quasi elastic form factors to describe the 
$\Delta$ excitation and consistency of the approach was investigated in \cite{marteau_delta}.

More details about the model can be found  in \cite{sobczyk_nuint02, marteau}.
The model includes the RPA and $\Delta$ in-medium self-energy effects. Recently, the model was upgraded and it 
now contains LDA effects. The elementary 2p-2h responses were added following the procedures described in \cite{marteau_thesis}.
The model in this paper -- like the MEChM model -- does not include RPA corrections to the np-nh
contribution (exceptions are terms describing 
pionless $\Delta$ decays). The elementary response functions used in the microscopic model 
are similar to those used in the MEChM paper, as shown in Fig. 3 in \cite{martini1}.

In Appendix A we collect basic formulae which allow for a comparison with other approaches.

\begin{figure}
\includegraphics[width = \columnwidth]{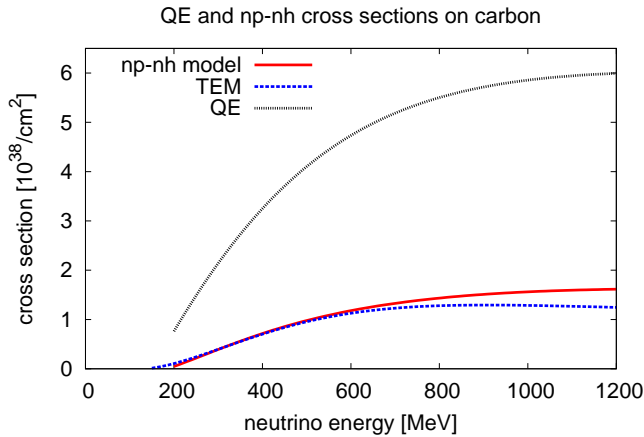}
\caption{\label{fig:after_rpa}  (Color online) Predictions for the np-nh contributions to the CC scattering on carbon from two models
discussed in this paper.  For comparison the QE cross section is also shown.}
\end{figure}

\begin{figure}
\includegraphics[width = \columnwidth]{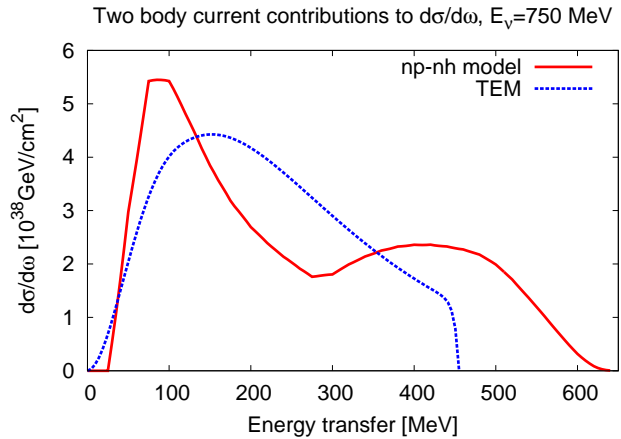}
\caption{\label{fig:diff_700} (Color online) Predictions for the two-body current contribution to the 
CC carbon target differential cross section versus energy transfer from the two models 
discussed in this paper.}
\end{figure}

In the microscopic model there is the possibility of two- and three- nucleon ejection. 
The three nucleon ejection contribution comes from 
pionless $\Delta$ decays i.e. from reactions $N\Delta \rightarrow NN$ and $NN\Delta \rightarrow NNN$ with $N$ standing for a nucleon 
\cite{oset_salcedo}. 
%
\subsubsection{Transverse Enhancement Model}
%
A new approach to describe CCQE scattering on nuclear targets is proposed in~\cite{bodek}. The model
is easy to implement in MC event generators. 
It is sufficient to modify the vector magnetic form factor keeping all other ingredients of the CCQE model as in the free nucleon target case. 

The authors of \cite{bodek} proposed for the carbon target a universal transverse enhancement function of $Q^2$. 
For low $Q^2$ its form is determined by scaling arguments while for large $Q^2$ ($>0.5$~GeV$^2$) it is obtained as a 
fit to
inclusive electron cross section data from the JUPITER experiment. 
The prescription to include transverse enhancement contribution in the numerical computations amounts to the replacement:
\begin{equation}
G_M^{p,n}(Q^2)\rightarrow \tilde G_M^{p,n}(Q^2)=\sqrt{ 1 + AQ^2 \exp (-\frac{Q^2}{B}) } G_M^{p,n}(Q^2)
\end{equation}
where $G_M^{p,n}(Q^2)$ are electromagnetic form-factors, $A= 6$ GeV$^{-2}$ and $B=0.34$~GeV$^{2}$.

The most interesting feature of the TEM model is that it offers a possible explanation to the 
apparent contradiction between low (MB) and high (NOMAD) 
neutrino energy $M_A$ measurements: for energies up to $\sim 700$~MeV the TEM predicts the
cross section to be similar to CCQE with $M_A=1.3$~GeV.
For higher neutrino energies the TEM cross section becomes significantly smaller and at $E_\nu \sim 5$~GeV it corresponds to CCQE with 
$M_A\sim 1.15$~GeV.

As the TEM prediction for the two-body current contribution one takes the difference between the cross sections calculated with the modified
and default magnetic form factors:
\begin{equation}
\frac{d^2\sigma^{TEM}}{dqd\omega} = \frac{d^2\sigma^{CCQE}}{dqd\omega} (\tilde G_M^{p,n}) - \frac{d^2\sigma^{CCQE}}{dqd\omega} (G_M^{p,n}).
\end{equation}
In the TEM only two-nucleon ejection takes place.
%
\subsection{Muon CC models comparison}
%
Fig. \ref{fig:after_rpa} shows the predictions for the np-nh contribution from both models. 
They are quite similar in size and only for neutrino energies above $700$~MeV does 
the microscopic model
predict a larger cross section. For comparison, the predictions for the QE cross section (one outgoing nucleon at the interaction point) 
are shown as well.
In the presented models, the multinucleon ejection contribution amounts to about 25\% of the QE cross section.

Fig.  \ref{fig:diff_700} shows the differential cross section for the multinucleon ejection contribution
as a function of the energy transfer. It should be stressed that so far no assumptions about the final state nucleons
were necessary. The model predictions are quite different. In the case of the microscopic model there is 
a lot of structure
coming from various ingredients of the model: contributions from pionless $\Delta$ decays, elementary $NN$ and $N\Delta$ 2p-2h
responses. Similar structure can be seen also in Fig. 2 in \cite{martini1}. 
There are two versions of the MEChM model which differ by elementary 2p-2h response functions.
Following the procedures described in \cite{martini1}, by making a fit to the data contained in \cite{aem}
new responses as functions of the variable 
$x \equiv Q^2/2M\omega = (q^2-\omega^2)/2M\omega$ (where $M$ is the nucleon mass) 
were obtained.
However, within the microscopic model of this paper the new responses lead to a too rapidly increasing np-nh cross section 
as a function of neutrino energy and in the rest of this paper the original Marteau responses are used \cite{marteau_thesis}. 

In the case of the TE model a sharp fall in the differential cross section at $\sim 460$~MeV comes from the way in which the model was implemented.
We followed the original paper and assumed the target nucleon to be at rest. Pauli blocking effects are introduced by means of a $Q^2$ 
dependent suppression function. An alternative implementation is to use the Fermi Gas model, but it is not obvious which approach
leads to better agreement with the MB 2D CCQE data \cite{sobczyk_tem}.

\begin{figure}
\includegraphics[width = \columnwidth]{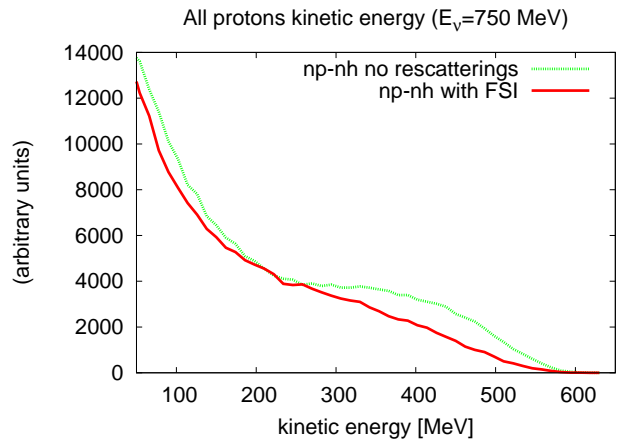}
\caption{\label{fig:kin_800} (Color online) Total kinetic energy of the final state protons coming from two-body current interactions 
as modeled by the microscopic model
implemented in NuWro. Predictions from the model with and without nucleon rescatterings are compared. The neutrino energy is $750$~MeV. }
\end{figure}

\begin{figure}
\includegraphics[width = \columnwidth]{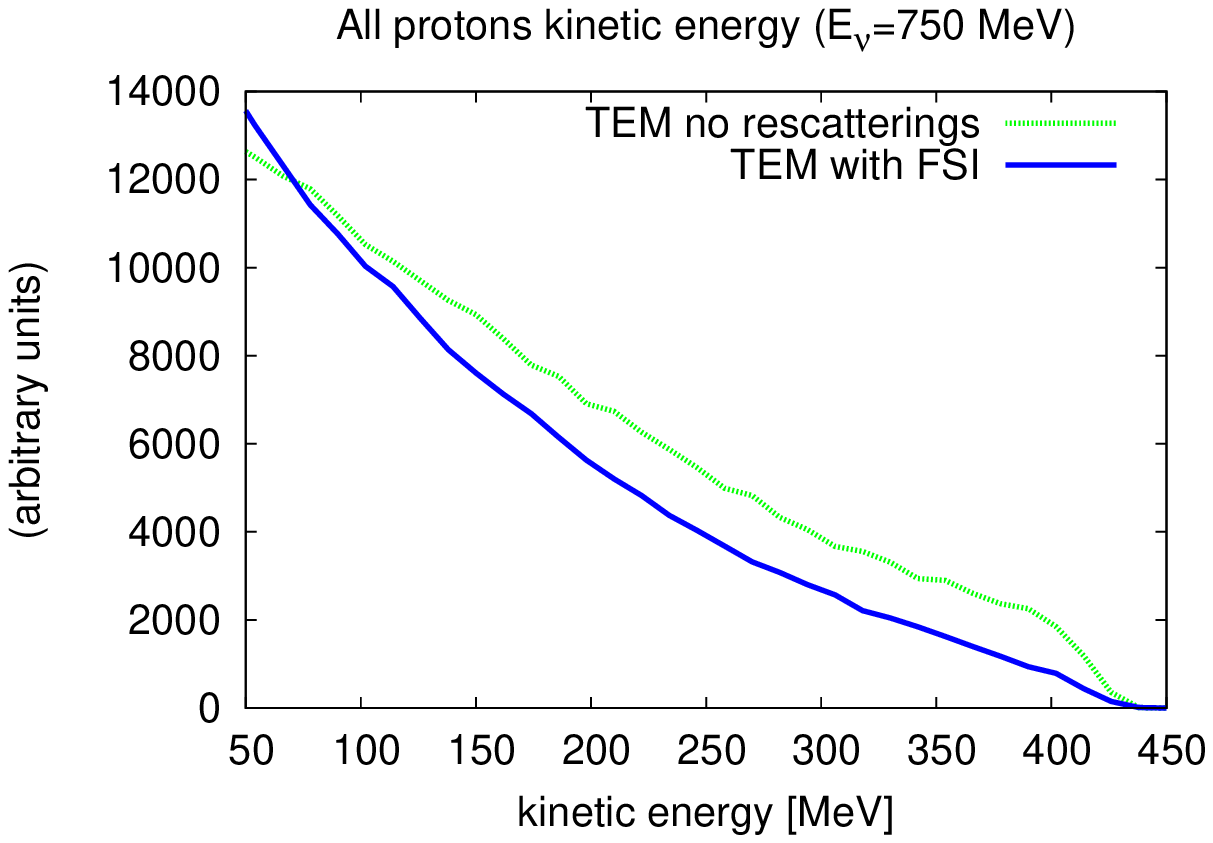}
\caption{\label{fig:kin2_800} (Color online) The same as in Fig. \ref{fig:kin_800} but for the TE model. }
\end{figure}

\section{Results}

\begin{figure}
\includegraphics[width = \columnwidth]{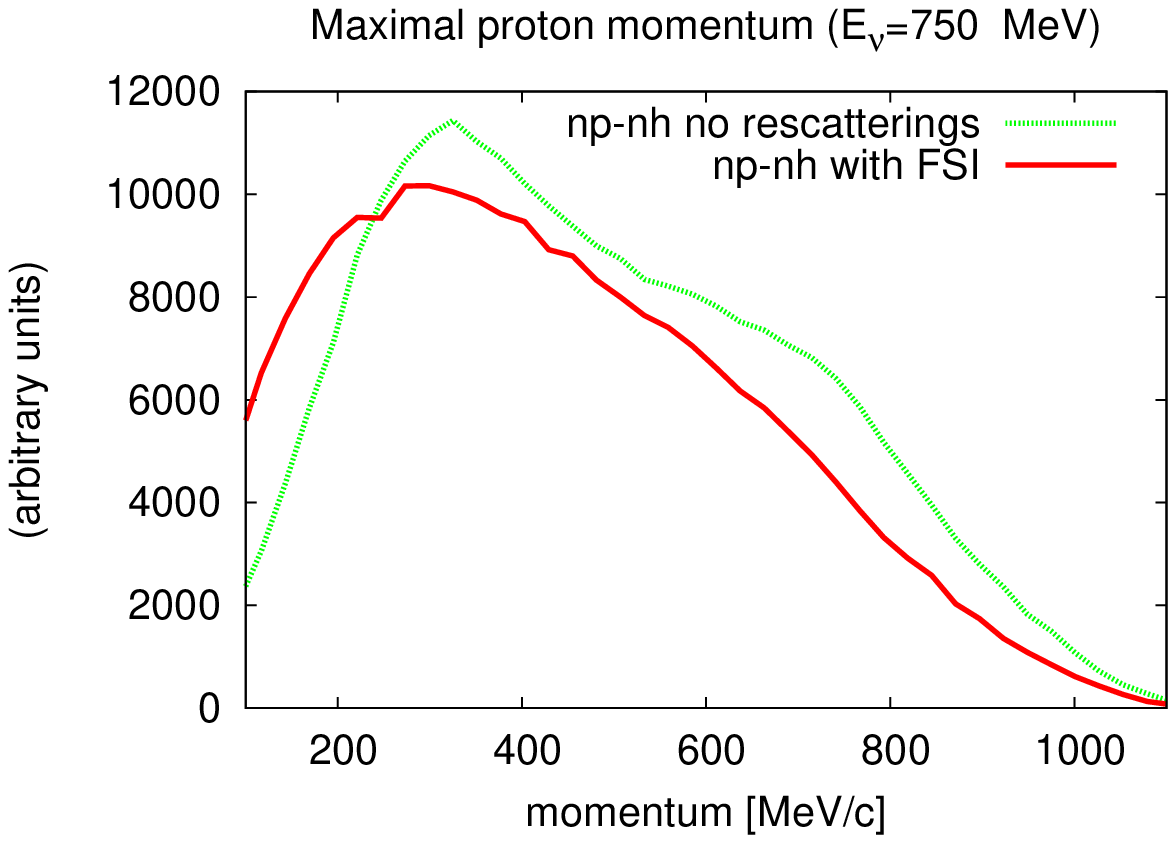}
\caption{\label{fig:max_800} (Color online) Momentum of most energetic final state protons coming from two-body current interactions 
as modeled by the microscopic model
implemented in NuWro. Predictions from the model with and without nucleon rescatterings are compared. The neutrino energy is $750$~MeV.
}
\end{figure}

\begin{figure}
\includegraphics[width = \columnwidth]{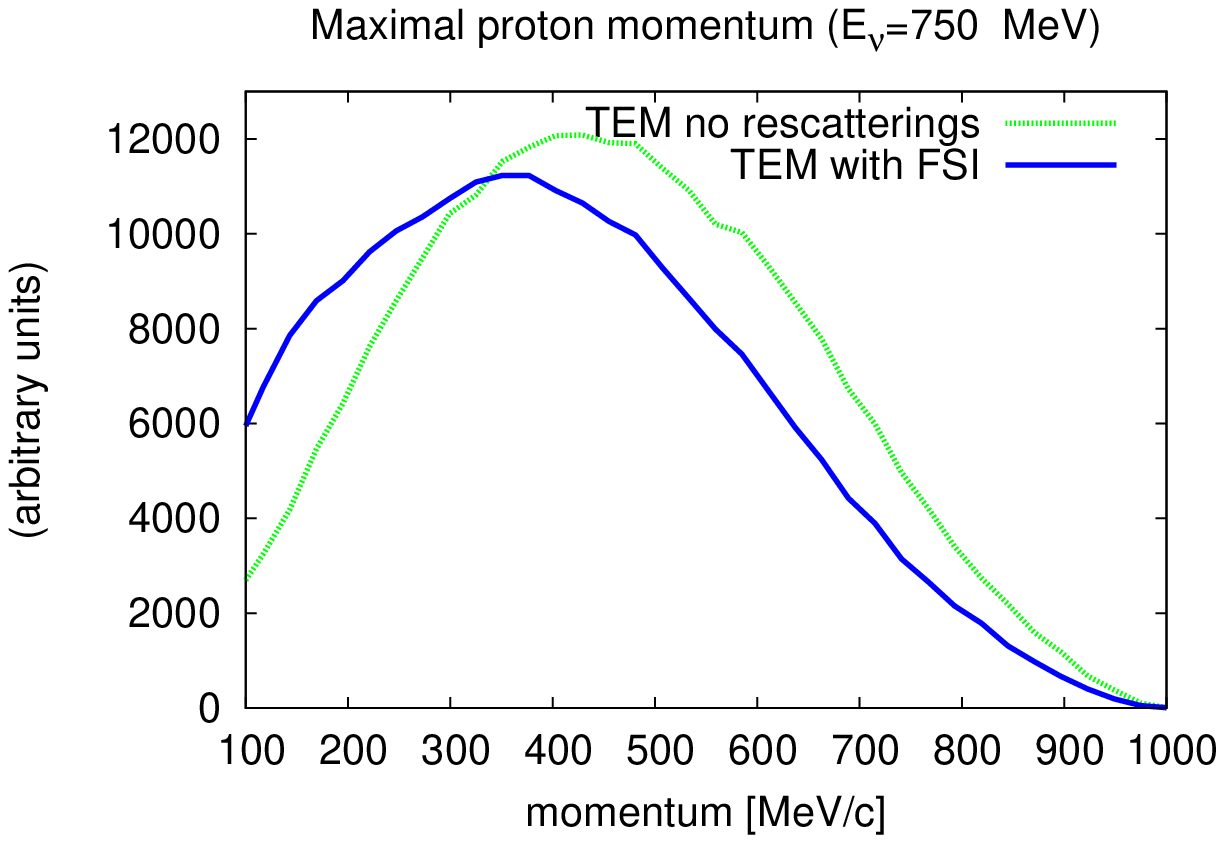}
\caption{\label{fig:max2_800} (Color online) The same as in Fig. \ref{fig:max_800} but for the TE model.
}
\end{figure}

\begin{figure}
\includegraphics[width = \columnwidth]{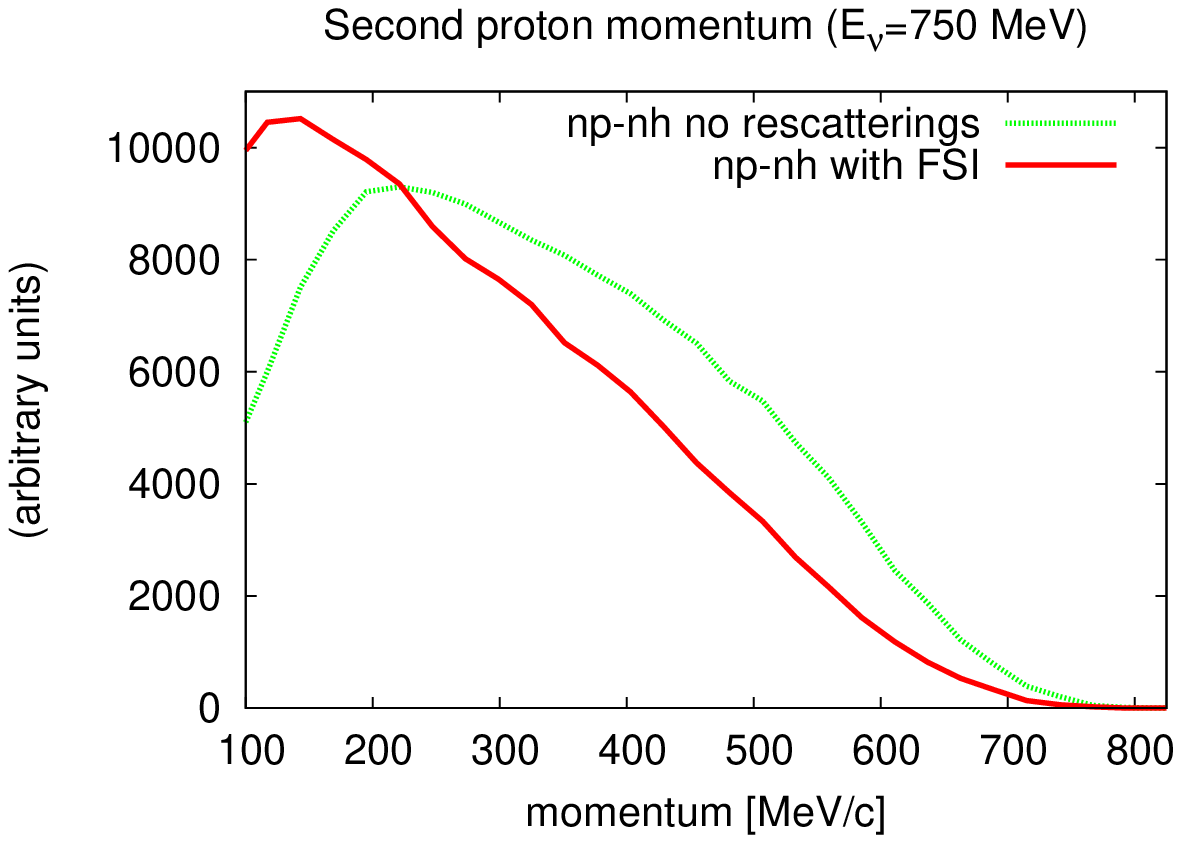}
\caption{\label{fig:second_800} (Color online) Momentum of second most 
energetic final state protons coming from two-body current interactions 
as modeled by the microscopic model
implemented in NuWro. Predictions from the model with and without nucleon rescatterings are compared. The 
neutrino energy is $750$~MeV.
}
\end{figure}

\begin{figure}
\includegraphics[width = \columnwidth]{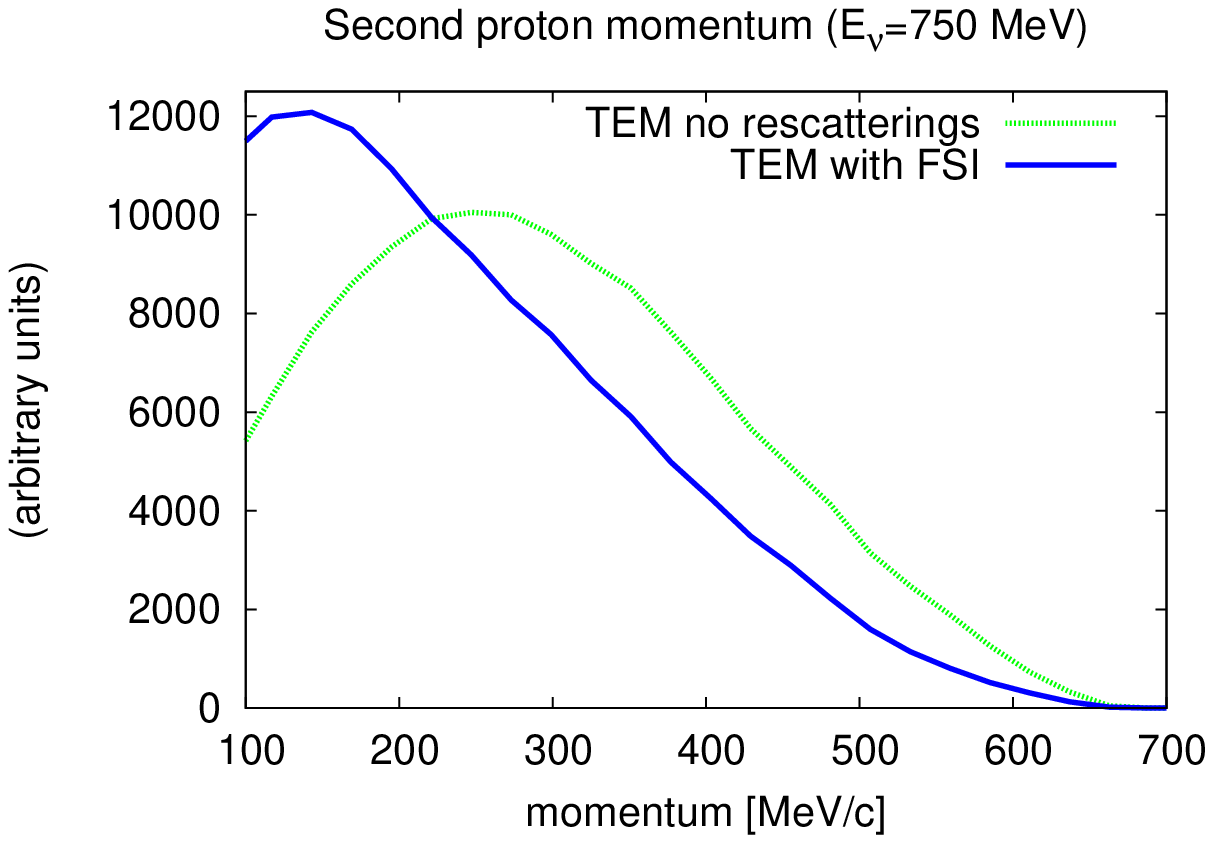}
\caption{\label{fig:second2_800} (Color online) The same as in Fig. \ref{fig:second_800} but for the TE model.
}
\end{figure}

\begin{figure}
\includegraphics[width = \columnwidth]{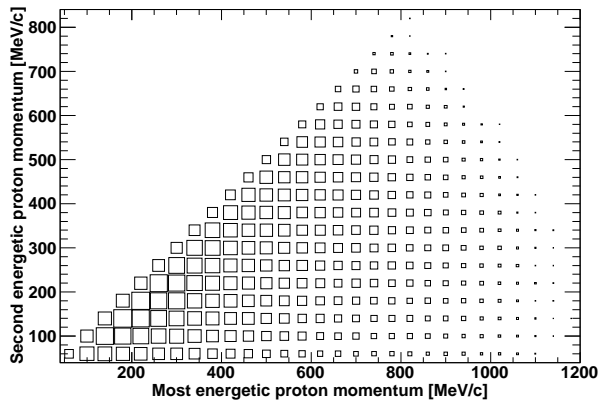}
\caption{\label{D2_bodek} Two dimensional distribution of proton momenta: highest energy with respect to the 
second highest energy protons.
Muon neutrino energy is $750$~MeV, simulations are done for the microscopic model.
}
\end{figure}

\begin{figure}
\includegraphics[width = \columnwidth]{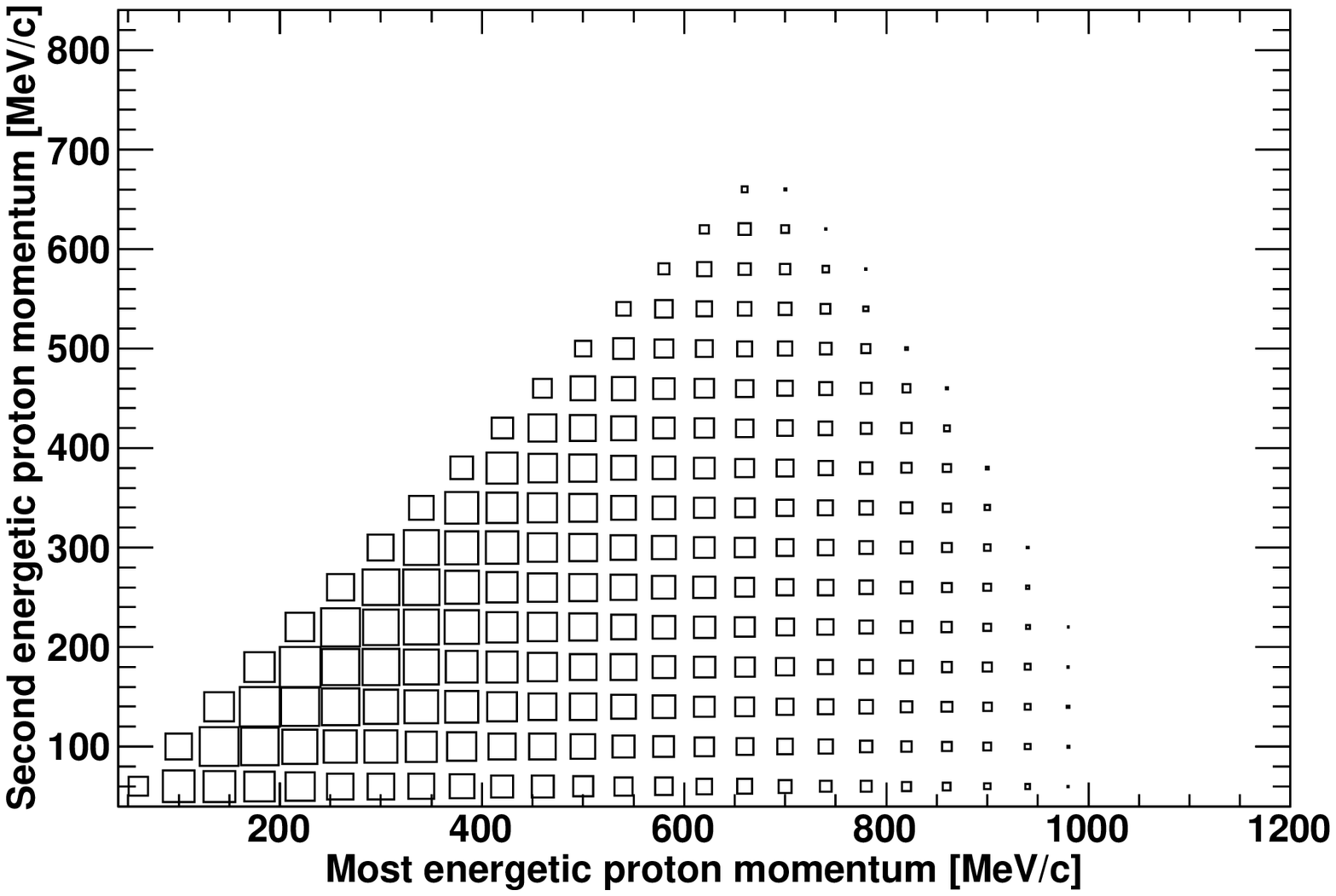}
\caption{\label{D2_npnh} 
The same as in Fig. \ref{D2_npnh} but for the TE model.
}
\end{figure}

Figures \ref{fig:kin_800} -- \ref{fig:second2_800} show predictions from both models
implemented in the NuWro MC event generator. All of them are for protons only
because in the experimental analysis neutrons are usually not detected. In Section III we focus on the impact of FSI effects
on the results.
%
\subsection{Total kinetic energy}
%
Figs \ref{fig:kin_800} and \ref{fig:kin2_800} show the kinetic energy of all the protons in the final state. 
Due to rescatterings protons lose a fraction
of their kinetic energy. No protons in the final state is a possible outcome in the NuWro FSI model.
%
\subsection{Nucleon momenta}
%
Figs \ref{fig:max_800} -- \ref{fig:second2_800}
show the distributions of the momenta of highest energy and second highest energy protons in the final state. 
Again, because of reinteractions protons becomes less energetic. 
On average their momenta are
reduced by $\sim 100$~MeV/c. 

An interesting feature of the distribution of the second most energetic protons is that the rescatterings make
the fraction of events in which there is at most one
proton in the final state smaller. 

The distributions seen in Figs \ref{fig:second_800} and \ref{fig:second2_800} are important
because they allow for an estimate of the probability that there is a pair of reconstructed protons in the final state. 
It is interesting that at the neutrino energy of only $750$~MeV the second nucleon can be quite energetic.

Figs \ref{D2_npnh} -- \ref{D2_bodek} show two-dimensional distributions: highest energy versus second highest energy protons for both models.
Because the range of energy transfers for the TEM is smaller, the phase space covered by pairs of protons 
is also reduced.
%
\subsection{Energy reconstruction}

\begin{figure}
\includegraphics[width = \columnwidth]{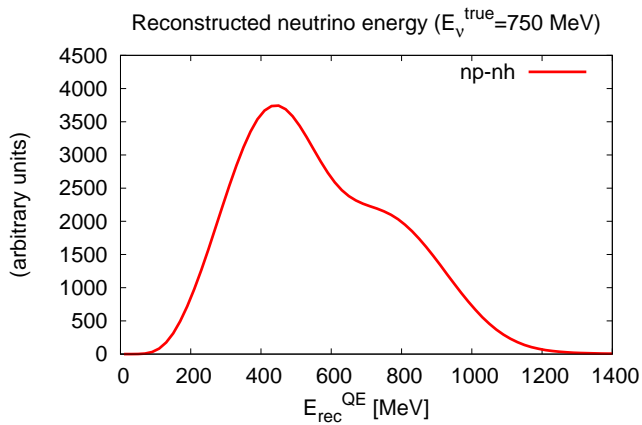}
\caption{\label{fig:reconstr_800} (Color online) Reconstructed neutrino energy for two 
body current interactions of $E_\nu = 750$~MeV muon neutrino.}
\end{figure}

Fig.  \ref{fig:reconstr_800} shows the 
distribution of reconstructed energy $E_{rec}^{QE}$ obtained within the microscopic model. 
The true neutrino energy
is always $750$~MeV. $E_{rec}^{QE}$ is defined based on the information infered from the final muon only, assuming that the interaction
is CCQE and the target nucleon is at rest:
\begin{equation}
E_{rec}^{QE} = \frac{2ME_\mu-m^2}{2(M-E_\mu + \sqrt{E_\mu^2 -m^2}\cos\theta_\mu)}
\end{equation}
where $E_\mu$ is muon energy, $m$ is muon mass, and $\theta_\mu$ is an angle between neutrino and muon three momenta.

The presence of the two-body current contribution introduces an important bias and
it can be seen that on average the true neutrino energy is larger by $\sim 150-200$~MeV than $E_{rec}^{QE}$. 

The TEM uses the QE kinematics and it is not suitable to study the bias in energy reconstruction.
%
\section{Discussion}
%
\subsection{A role of energy transfer}
%
The predictions from both models implemented in NuWro are quite similar in shape; the only major difference
is that in the microscopic model protons are on average more energetic. 
This can be understood as a consequence of different shapes for the muon differential cross
sections in energy transfer. In fact, this energy (a fraction of) is seen as the kinetic energy of ejected protons. 
From the point of view of the multinucleon ejection model, 
the most important features of muon CC two-body current contribution to the differential cross section are: 
(i) the integrated size, and (ii)
the distribution of events in the energy transfer. 
%
\subsection{How to see two-body current events}
%
In the $1$~GeV energy region (e.g. as it is in the T2K experiment) two-body current events are almost all CCQE-like. 
The probability to produce pions due to FSI effects 
is negligible.  It follows that in the experimental analysis it is important to 
develop an effective veto on pions. 

There are two promising observables which contain an information about the two-body current contribution. 
The first one is pairs of protons in the final state
both with momenta above some threshold value. The second one is the integrated kinetic energy of charged hadrons in the final state. 
In fact, if the typical neutrinos energies are about or below $1$~GeV,
most of the protons will most likely remain  not reconstructed and they will only contribute to the vertex activity. 

All the computations discussed below are done for the $750$~MeV muon neutrinos. They can give an idea about the size of effect
to be expected e.g. in the T2K experiment. The computations are done for the models as implemented in NuWro with
all the FSI effects being included. 
%
\subsubsection{Proton pairs}
%
Tables \ref{tabelka_tem} and \ref{tabelka_npnh} show the predicted number of proton pairs with both 
momenta above various threshold values. The signal is defined as exactly two protons 
above: $300$, $400$ and $500$~MeV/c. It is assumed that in the samples of events there are no $\pi^0$ ($\pi^0$ is either 
reconstructed or at least
one energetic photon resulting from its decay is detected). 
As for $\pi^\pm$ it is assumed that they can be identified if their momenta are above either $0$ (i.e. all of them are 
detected and the sample of events contains no $\pi^\pm$), or $200$~MeV/c. 
In each situation the numbers of signal and background events are shown separately. The background events are those with exactly two protons
satisfying the selection criteria but originating from other dynamical mechanisms. 
%
\begin{table}
\begin{tabular}{c||l|l|l|l|}
$\pi^\pm$ cut [$\frac{MeV}{c}$]$\downarrow$& proton cut [$\frac{MeV}{c}$] $\rightarrow$ & 300 & 400 & 500 \\
\hline\hline
0 &                  signal & 7185     & 4201  &  1805  \\
               & background & 13774     & 7928 &  2311 \\
\hline
200 & signal & 7231      & 4201  &  1805  \\
 & background & 16158     & 8577 &  2388 \\
\hline
\end{tabular}
\caption{Predicted number of proton pairs with both momenta above various threshold values and two threshold values of the $\pi^\pm$
momentum. 
Simulations were done for $750$MeV muon neutrinos. The number of generated events is $2.5\cdot 10^5$. 
The microscopic model in a NuWro implementation was used to 
produce signal events.
}
\label{tabelka_npnh}
\end{table}
%
\begin{table}
\begin{tabular}{c||l|l|l|l|}
$\pi^\pm$ cut [$\frac{MeV}{c}$]$\downarrow$& proton cut [$\frac{MeV}{c}$] $\rightarrow$ & 300 & 400 & 500 \\
\hline\hline
0 &                  signal & 5457     & 2271  &  651  \\
               & background & 13780     & 7961 &  2267 \\
\hline
200 & signal & 5465      & 2271  &  651  \\
 & background & 16112     & 8691 &  2349 \\
\hline
\end{tabular}
\caption{
The same as in TABLE \ref{tabelka_tem} but for the NuWro implementation of the TE model.
}
\label{tabelka_tem}
\end{table}
%
The total number of generated CC events is $2.5\cdot 10^5$. In the case of microscopic model for multinucleon ejection
the sample of events contains: $59.4$\% QE and $15.2$\% two-body current events. Remaining are 
mostly single pion production events. For the TEM the composition of the sample of events is similar.
The background for the two proton signal comes mainly from the real pion production and its subsequent 
absorption. The NuWro implementation of the absorption contains only two-body mechanism, and the size of the background can be 
overestimated
\cite{lads}.

There are noticable differences in the predictions for the number of the signal events from the two analyzed models,
and they show an interesting pattern.
In the case of pairs of protons both with momenta above $500$~MeV/c the difference is by a factor of 3 while for protons with
momenta above $400$~MeV/c and $300$~MeV/c the differences are only $\sim 80$\% and  $\sim 30$\%. 
For larger proton momenta
the pattern can be explained by the fact that the probability of a large energy transfer is in the case of microscopic model much bigger.
For lower proton momenta the most important becomes the fact that overall cross 
sections for multinucleon ejection are in both models
quite similar. 

The numbers of background events are slightly different in both cases. They were generated in two separate simulations and also
the cross sections for multinucleon ejection are slightly different. 

The numbers shown in Tables \ref{tabelka_tem} and \ref{tabelka_npnh} 
suggest that in order to separate the signal from the background it is enough to have the precision in the 
background estimation on the level of $50$\%, which seems to be a realistic goal to achieve. 

We conclude that counting of proton pairs can provide an experimental proof of existence of the multinucleon ejection contribution
and also a tool to discriminate between various models. However, it is necessary to have large samples of event because the probability to 
have a signal pair of protons with both momenta above $500$~MeV/c is only $0.26-0.72$\%.
%
\subsubsection{Integrated energy of charged hadrons}
%

\begin{figure}
\includegraphics[width = \columnwidth]{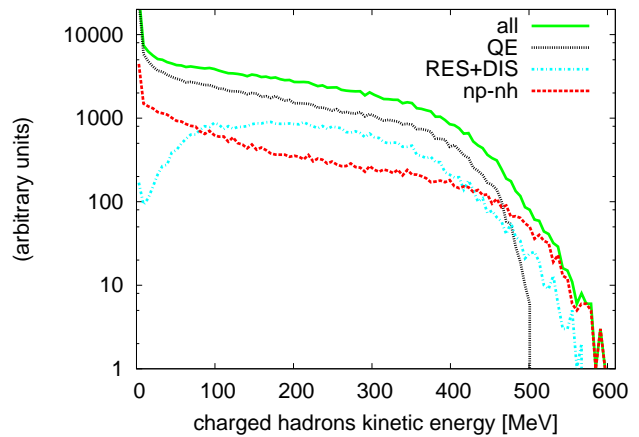}
\caption{\label{fig:deposit} (Color online) The distribution of hadronic kinetic energy for the events with cuts as described in the text.
Simulations were done with $750$~MeV muon neutrinos using NuWro implementation of the microscopic model.
The number of generated events is $2.5\cdot 10^5$. Contributions from CCQE, RES+DIS and np-nh events are also shown separately.
}
\end{figure}

\begin{figure}
\includegraphics[width = \columnwidth]{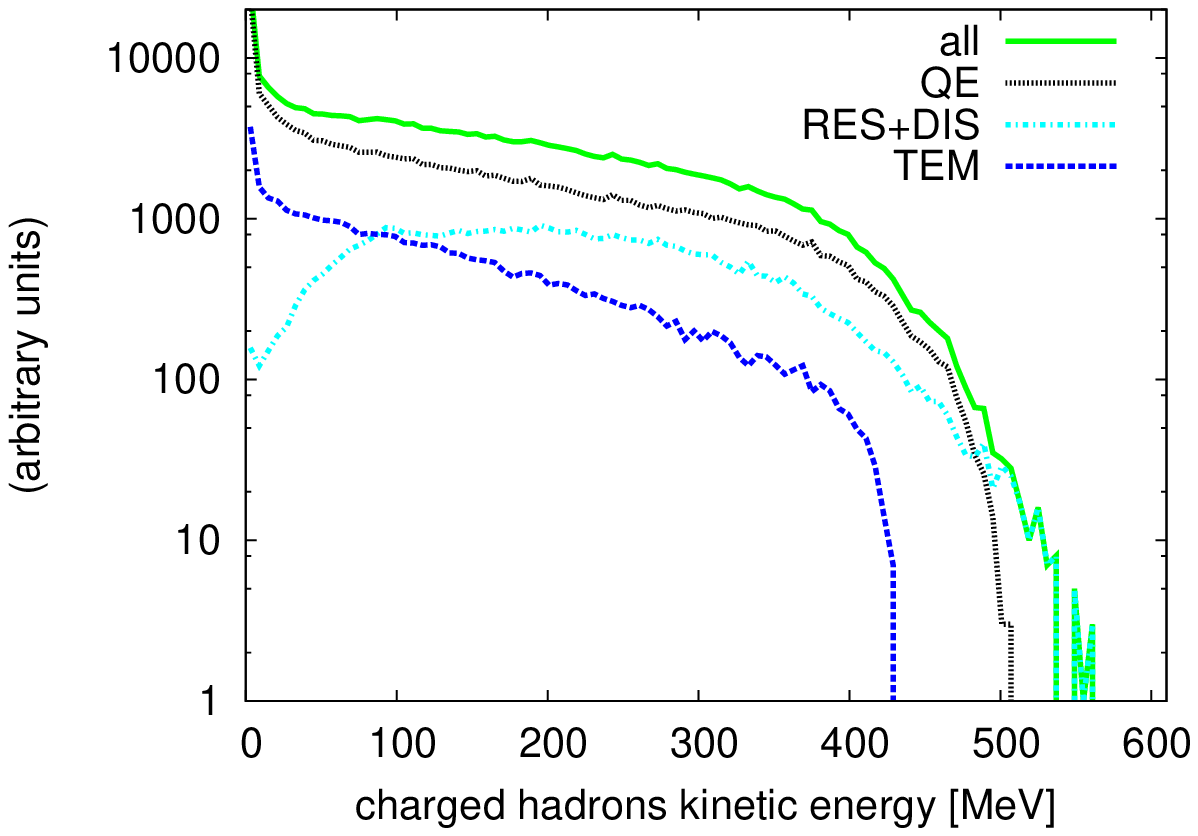}
\caption{\label{fig:deposit2} (Color online) The same as in Fig. \ref{fig:deposit} but for the TE model.
}
\end{figure}

\begin{figure}
\includegraphics[width = \columnwidth]{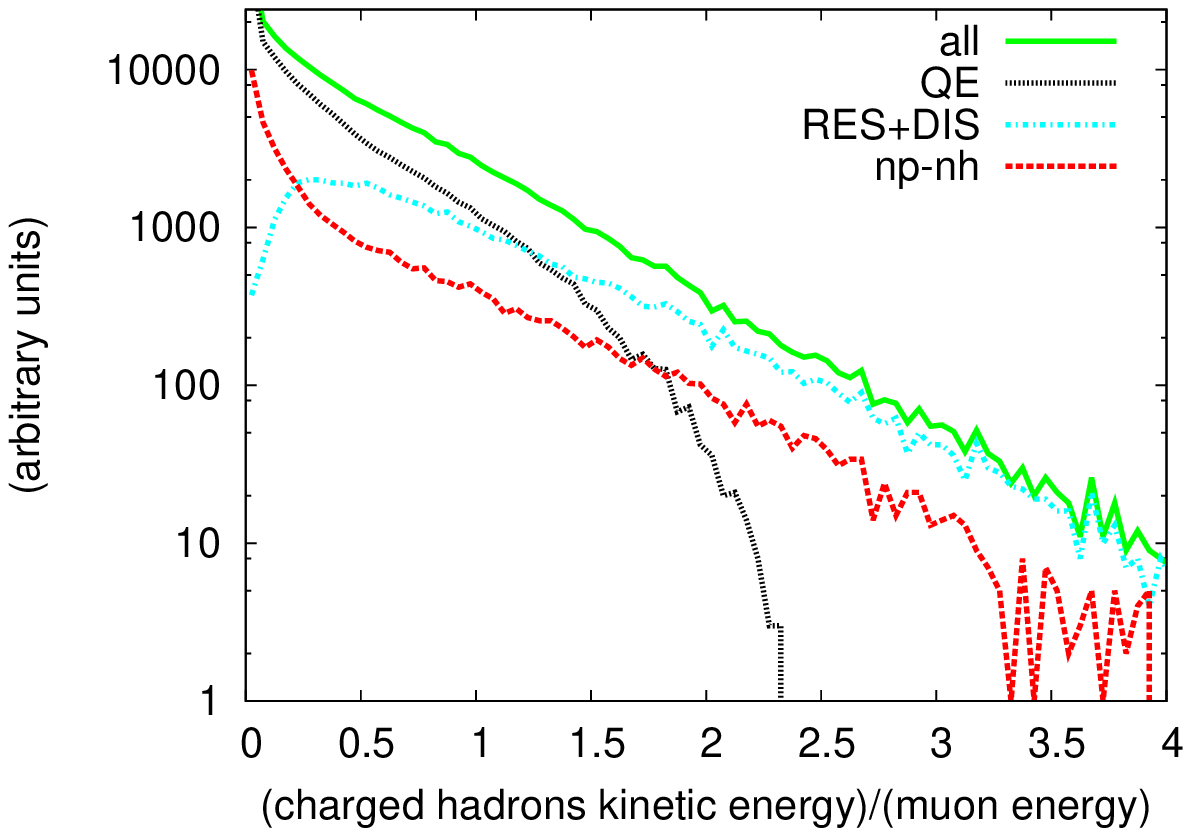}
\caption{\label{fig:ratio} (Color online) The distribution of hadronic kinetic energy normalized by the muon energy (see Eq. \ref{hadrons}) 
for the events with cuts as described in the text. Simulations
were done with $750$~MeV muon neutrinos using NuWro implementation of the microscopic model.
The number of generated events is $2.5\cdot 10^5$. Contributions from CCQE, RES+DIS and np-nh events are also shown separately.}
\end{figure}

\begin{figure}
\includegraphics[width = \columnwidth]{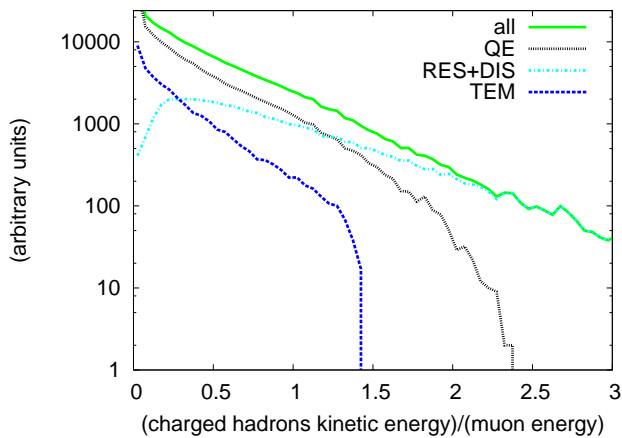}
\caption{\label{fig:ratio2} (Color online) The same as in Fig. \ref{fig:ratio} but for the TE model.
}
\end{figure}

For the overall charged hadron kinetic energy two observables can be considered:
\begin{equation}
 \sum_j T_j, \qquad{\rm and}\qquad \frac{\sum_j T_j}{E_\mu}.
\label{hadrons}
\end{equation}
where $j$ runs over charged hadrons in the final state and $E_\mu$ is the muon energy. 
All the protons contribute to the observables defined above, independent on how large their momenta are.
Also in this case one must 
introduce some assumptions about 
the pions. In the plots shown in Figs \ref{fig:deposit} -- \ref{fig:ratio2} in the signal events there are no $\pi^0$ and no $\pi^\pm$
with momenta above $200$~MeV/c. In the figures contributions from CCQE, RES+DIS and two-body current mechanism are also shown
separately.  In the neutrino MC nomenclature
RES refers to the resonance region and DIS to more inelastic events but for the purpose of this study they are combined together.
All three contributions are different in shape. This is clearly seen in the case of RES+DIS. The shape
of the multinucleon ejection contribution depend on the underlying muon cross section model and on the FSI effects. 
It is important to remember that in the
distributions shown in Figs \ref{fig:deposit} and \ref{fig:ratio} also reconstructed protons are included. 

In the actual experimental analysis the directly observable quantity is the reconstructed energy. However, using neutrino MC generator one
can make predictions for the shape of the distribution of the reconstructed energy and compare to the data. It can happen that both measured
and predicted shapes do not match and that the inclusion of the contribution from the two-body current mechanism
one gets much better data/MC agreement.
%
\section{Final remarks}
%
The aim of this paper was to discuss the phenomenological consequences of the two-body current contribution to
the neutrino CC muon inclusive cross section. In order to estimate the size of effect for the hadron observables 
the multinucleon ejection model 
was proposed which can be easily implemented in neutrino MC event generators. Using the model
some 
observables which contain an information about two-body current dynamics were discussed.

It is important that the multinucleon ejection model proposed in the Section \ref{model} can be used together with any model
providing predictions for the two-body contribution to the CC inclusive neutrino-nucleus cross section. 

Identification of the two-body current experimental signal is possible if based on a careful data/MC comparison. For that
it is essential to have a reliable model of FSI effects. It is also central to have a good description of the
nucleon final states arising after pion absorption. 

One should also keep in mind that on the theoretical side there is a lot of uncertainty as to how large the two-body current 
contribution is expected to be. A careful comparison of the predictions from the models described in the Introduction for
the MiniBooNE flux averaged contribution in the region $\cos\theta_\mu\in (0.8, 0.9)$ (see Fig. 6 in \cite{martini2} , Fig. 3 in 
\cite{nieves2} and Fig. 4 in \cite{barbaro}) reveals that they differ by a factor of two. 

All the microscopic models discussed in this paper are based on the Local Fermi Gas model and the arguments of the paper
\cite{carlson} indicate that in order to evaluate the two-body current contribution it is necessary to use a more realistic 
model of the nucleus ground state. For the multinucleon ejection model this means that one should include
more correlations between initial state nucleons. 

During the work on the upgraded version of this paper an article on the similar subject was
published by Lalakulich et al \cite{olga}.
%
\appendix
%
\section{Some formulas from the microscopic model used in this paper}
%
The components of the hadronic tensor are expressed as:
\begin{equation}
 H^{\mu\nu} = H^{\mu\nu}_{NN} + H^{\mu\nu}_{N\Delta}+ H^{\mu\nu}_{\Delta N}+ H^{\mu\nu}_{\Delta\Delta}.
\end{equation}
For example,
\[ H^{00}_{XY} = \sqrt{\frac{M_X + \sqrt{M_X^2 + q^2}}{2M_X} } 
\sqrt{\frac{M_Y + \sqrt{M_Y^2 + q^2}}{2M_Y} }\]
\begin{equation}
 \left( \alpha_{0X}^0\alpha_{0Y}^0 R^c_{XY} + \beta_{0X}^0\beta_{0Y}^0 R^l_{XY}\right),
\end{equation}
\[
 \alpha_{0X}^0 = F_1(\omega ,q) - F_2(\omega ,q) \frac{q^2}{2M(M_X+E^X_q)}
\]
\begin{equation}
\beta_{0X}^0 = \frac{q}{M_X+E^X_q}\left( G_A(\omega ,q)  - \frac{\omega}{2M} G_p(\omega ,q)\right).
\end{equation}
$F_1$, $F_2$, $G_A$ and $G_p$ are standard form factor in the weak nucleon-nucleon transition matrix element.

In the case of free Fermi gas:
\[R_{N\Delta} = R_{\Delta N} =0,\]
\begin{equation}
 R_{NN}^{c,l,t} (\omega , q) = \Im (\Pi^0_{NN} )(\omega , q),
\end{equation}
\begin{equation}
 R_{\Delta\Delta}^{l,t}(\omega , q) = \Im (\Pi^0_{\Delta\Delta}) (\omega , q) ,
\end{equation}
\[\Im  (\Pi^0_{NN})(\omega , q) = -\frac{M^2}{2\pi^2}\]
\[
 \int d^3p \frac{\delta(\omega + E^N_{\vec p} - E^N_{\vec p+\vec q})}{E^N_{\vec p}E^N_{\vec p+\vec q}} \Theta(k_F-|\vec p|)
\Theta (|\vec p+\vec q|-k_F)=
\]
\begin{equation}
=-\frac{M^2}{\pi q}(E_1-E_2)\Theta(E_1-E_2),
\end{equation}
where
$E_1=E_F$, $E_2=max(-\frac{\omega}{2}, E_F-\omega , \frac{q}{2}\sqrt{1+\frac{4M^2}{Q^2}})$.
\begin{equation}
 \Im (\Pi^0_{\Delta\Delta})(\omega , q) = - \frac{4M_\Delta^2}{9\pi^3}
\int d^3p \frac{\Gamma_\Delta \Theta (k_F-|\vec p|)}{(s-M_\Delta^2)^2 + M_\Delta^2\Gamma_\Delta^2}.
\end{equation}
$s=(p+q)^2$.

In the nuclear matter $\Gamma_\Delta$ is modified because of the Pauli blocking effect which we introduce as a multiplicative factor
$P_B$ and also by the $\Delta$ self energy:
\[\Gamma_\Delta \rightarrow P_B\cdot\Gamma_\Delta -2\Im (\Sigma ).\]
Our parameterization of $\Im (\Sigma )$ is based on \cite{oset_salcedo}:
\[ \Im (\Sigma ) = \Im (\Sigma^{N\pi} ) + \Im (\Sigma^{NN} ) + \Im (\Sigma^{NNN} )\]
where the last two terms correspond to pionless $\Delta$ decays with an emission of two and three nucleons. We converted the information
contained in \cite{oset_salcedo} to the kinematical situation of the lepton scattering using the procedure described in \cite{sobczyk_nuint02}.
Because $\Im (\Sigma )$ depends strongly on the nuclear density it was evaluated (and tabularized) in 
the Local Density Approximation using the density profile of the carbon nucleus:
\[
 \Im (\Pi^0_{\Delta\Delta})_{LDA}(\omega , q) = 
\]
\begin{equation}
- \frac{4M_\Delta^2}{9\pi^3}\int d^3r \rho(\vec r)
\int d^3p \frac{\Gamma_\Delta (\vec r)\Theta (k_F-|\vec p|)}{(s-M_\Delta^2)^2 + M_\Delta^2\Gamma_\Delta^2 (\vec r)}.
\end{equation}
For the real part of the $\Delta$ self energy the parametrization 
\[M_\Delta \rightarrow M_\Delta + 40 MeV\frac{\rho}{\rho_0}\]
was used.

Details about the RPA computations can be found in \cite{sobczyk_nuint02}.


\begin{acknowledgments} I thank Marco Martini for explaining to me many details of the MEChM model.
I also thank Jorge Morfin for many discussions and comments on an earlier version of the manuscript of the paper,
and Steven Dytman, Luis Alvarez-Ruso and Gabe Perdue for several useful suggestions.
The author was
partially supported by the grants: N N202 368439 and DWM/57/T2K/2007.

\end{acknowledgments}


\end{document}